\documentclass[twocolumn,epsf,prb,amsmath,amssymb,showpacs]{revtex4}
\usepackage{graphicx}

\begin{document}
\title{Graphene-based resonant-tunneling strucures}
\author{J. Milton Pereira$^{1,2}$ Jr., P. Vasilopoulos$^3$, and F. M. Peeters$^1$}
\address{$^1$Department of Physics, University of Antwerp, Groenenborgerlaan 171, B-2020 Antwerpen, Belgium\\
$^2$Departamento de F\'{\i}sica, Universidade
Federal do Cear\'a, Fortaleza, Cear\'a, $60455$-$760$, Brazil\\
$^3$Department of Physics, Concordia University, 7141 Sherbrooke
Ouest, Montreal, Quebec, Canada H4B 1R6}

\begin{abstract}
Resonant electronic transmission through graphene-based double
barriers (wells) is studied as a function of the incident wave
vector, the widths and heights (depths) of the barriers (wells),
and the separation between them. Resonant features in the
transmission result from resonant electron states in the wells or
hole states in the barriers and strongly influence  the ballistic
conductance of the structures.

\end{abstract}
 \maketitle

Recently  graphene and graphene-based microstructures have been
realized experimentally. They exhibit  unusual properties and have
potential technological applications especially as an alternative
to the current Si-based technology. The relativistic-like
properties of carriers in graphene \cite{zheng,novo3,shara,zhang},
such as an unusual quantum Hall effect \cite{novo4},  result from
the gapless and approximately linear electron spectrum near the
Fermi energy at two inequivalent points of the Brillouin zone. The
charge carriers in these structures are described as massless,
chiral "relativistic" fermions, governed by the Dirac equation,
and their "relativistic" behavior is expected to lead to the
observation of perfect transmission across potential barriers,
known as Klein tunneling \cite{klein}. In addition, very recently
electronic confinement was demonstrated in graphene
microstructures using standard lithography methods \cite{berger}.
Recent studies have considered tunneling through {\it single}
graphene barriers \cite{kat}, wells \cite{milton}, and {\it n-p-n}
junctions \cite{falko}. One important aspect of the electronic
transport in quantum structures in graphene, that so far has not
been investigated, is resonance effects on the transmission. In this letter we
investigate them
as a function of the barrier
(well) height (depth) and separation, and  contrast those for {\it
double} barriers with those for "non relativistic" electrons.

The crystal structure of undoped  graphene layers is that of a
honeycomb lattice of covalent-bond carbon atoms. To each carbon
atom corresponds a valence electron and the structure is composed
of two sublattices, labelled A and B. The low-energy excitations
of the system in the vicinity of the ${\mathbf K}$ point and in
the presence of a potential $U$  are described by the 2D Dirac
equation,
\begin{equation}
\{v_F[\vec{\sigma}\cdot \hat{\mathbf p}]+m\, v_F^2 \sigma_z\} \Psi
= (E-U)\Psi,
\end{equation}
where the pseudospin matrix $\vec {\sigma}$ has components given
by Pauli's matrices, $\hat{\mathbf p} = (p_x,p_y)$ is the momentum
operator.  The "speed of light" of the system is $v_F$, i.e., the
Fermi velocity ($v_F \approx 1\times 10^6$ m/s). The eigenstates
of Eq. (1) are two-component spinors $\Psi = [\psi_A \, , \,
\psi_B]^T$, where $\psi_A$ and $\psi_B$ are the envelope functions
associated with the probability amplitudes at the respective
sublattice sites of the graphene sheet. The term $\propto m\,
v_F^2$ creates an energy gap and may arise due 
to an interaction with a substrate.

In the presence of a one-dimensional (1D) confining potential
$U=U(x)$ we attempt solutions of Eq. (1) in the form
$\psi_A(x,y)=\phi_A(x)e^{ik_y y}$ and
$\psi_B(x,y)=i\phi_B(x)e^{ik_y y}$ because of the translational
invariance along the $y$ direction. The resulting coupled,
first-order differential equations for $\phi_A(x)$ and $\phi_B(x)$
can be easily decoupled. They read
\begin{equation}
\frac{d^2 \phi_C}{d \xi^2} + (\Omega^2 - \beta^2)\phi_C -
\frac{u'}{\Omega_{\pm}}\left( \pm\frac{d \phi_C}{d \xi} -
\beta\phi_C \right) = 0.
\end{equation}
Here $\xi=x/L$, $\Omega_\pm = \epsilon - u\pm\Delta,$ $\Omega=
(\Omega_+\Omega_-)^{1/2},\ \beta=k_yL, \ u=UL/\hbar v_F, \
\epsilon = EL/\hbar v_F$, and $\Delta = m\,v_FL/\hbar$; $L$ is the
width of the structure, $u'$  the derivative of $u$ with respect
to $\xi$, and the $+ (-)$ sign refers to 
$C=A$ ($C=B$).

For a single quantum well, with a square potential of height
$U_0$, Eqs. (2)
admit solutions which describe electron
states confined across the well and propagating along it
\cite{milton}. For confined states, the spinor components decay
exponentially in the region $\xi < - 1/2$. Then   $\phi_A(x)$
can be written \cite{milton} as
\begin{eqnarray}
\nonumber \hspace*{-0.4cm}\phi_A(x)&=& e^{i\alpha \xi} + B_1
e^{-i\alpha \xi},\quad\quad  \xi < -1/2, \\* &=&A_2e^{i\kappa
\xi}+B_2e^{-i\kappa \xi}, \quad -1/2 \leq \xi \leq 1/2,\\*
 \nonumber
&=&A_3e^{i\alpha \xi},\quad\quad\quad\quad \quad\quad \xi > 1/2,
\end{eqnarray}
where  $\alpha =  [(\epsilon - u_0)^2-\beta^2 - \Delta^2]^{1/2}$
and  $\kappa^2 = \epsilon^2 - \beta^2-\Delta^2$.  Notice that
$\phi_A(x)$ depends on $k_y$ through $\beta=k_y/L$. A similar
expression holds for $\phi_B$. Then, the transmission coefficient
is obtained as $T = |A_3|^2$ and $A_3$ is determined by
matching $\phi_A$ and $\phi_B$ at $\xi=-1/2$ and $\xi=1/2$. This
procedure can be repeated for double wells
or barriers (with $\kappa\to -\kappa$) to obtain the
transmission $T$ through them. We avoid giving
any expressions for $T$ and instead concentrate on the results for it.
We obtain them without the mass term since $\Delta$ is usually
very
small.\\
{\it i) Double barriers}.
 A ($k_y, k_x$) contour plot of
the logarithm of the transmission $T$ through a double barrier is
shown in Fig. 1. Typical values for the barrier height $U_0 = 50$
meV, the barrier width   $L = 50$ nm, and the inter-barrier
separation $d=100$ nm were used. As seen, $T$ depends on the
direction of propagation or incident angle $\theta\
(\tan\theta=k_y/k_x$) and its overall behavior is similar to the
earlier single-barrier transmission \cite{kat} with the exception
of  well-defined resonances. 
The overall directional dependence of $T$ results
from the chiral
nature of the quasiparticles in graphene  and a resonance caused
by
confined hole states in the barrier. Notice the
perfect transmission $T=1$ for normal or near-normal incidence
($\tan\theta\approx 0$), which is a signature of Klein tunneling
\cite{kat}. Resonant transmission,
which is typical for ordinary
double-barrier structures \cite{fer}, is found for values of
$k_x$ that are integer multiples of $\pi/d$.

Another way of presenting the transmission
 is to plot $T$ versus the angle $\theta$ for fixed energy,  as shown in
Fig. 2 for $E=72, 78, 85$ meV by the red-dashed, black-solid, and
blue-dotted curve, respectively. As seen, the transmission is
perfect not only for normal incidence but also for particular
incidence angles, for which the values of the momentum components
match the resonance peaks shown in Fig. 1.

It is interesting to explore the transmission as a function of the
inter-barrier separation $d$. A ($d, k_y$) contour plot is shown in
Fig. 3 for $U_0 = 100$ meV, $E=27$ meV, and $L=50$ nm. Notice that
for small values of $k_y$, i.e. small incidence angles, the
transmission is perfect regardless of the magnitude of $d$. For
larger $k_y$ the transmission oscillates as a function of the
inter-barrier separation with a period that depends on the
momentum. This is a new feature,
{\it absent} from
ordinary resonant structures, in which $T$ is independent of $k_y$.

\begin{figure}
\vspace*{-1.2cm} \hspace*{-1cm}
\includegraphics*[height=6cm, width=8cm]{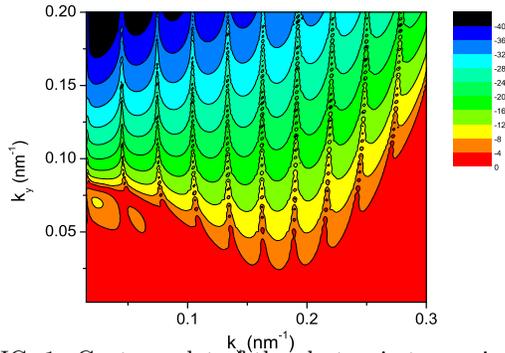}
\vspace*{-0.99cm} \caption{Contour plot of the electronic
transmission through a double barrier for $U_0 = 50$ meV, width $L
= 50$ nm, and inter-barrier separation $d=100$ nm. }
\end{figure}

{\it ii) Double wells}. The result for the transmission of
electrons, with energies above a single well, was given in Ref.
9
and that for a double well is shown in Fig. 4. Though some
features are qualitatively similar in both cases, that for a
double well is characterized by a strong enhancement of the
perfect transmission regions and has many resonant features that
are absent from that for a single well. As compared to the
double-barrier case, cf. Fig. 1, the resonances are much broader
and the transmission on the average much higher.

\begin{figure}
\vspace{-0.8cm} \hspace*{-1cm}
\includegraphics*[height=5cm, width=6cm]{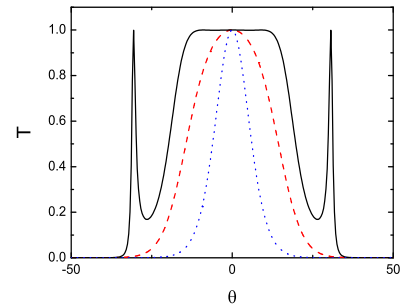}
\vspace*{-0.7cm} \caption{
Transmission through
 a double  barrier as a function of the  angle of incidence
for $U_0 = 100$ meV, $L = 50$ nm, and inter-barrier separation
$d=100$ nm. }
\end{figure}

\begin{figure}
\vspace*{-1.3cm} \hspace*{-1cm}
\includegraphics*[height=6cm, width=8cm]{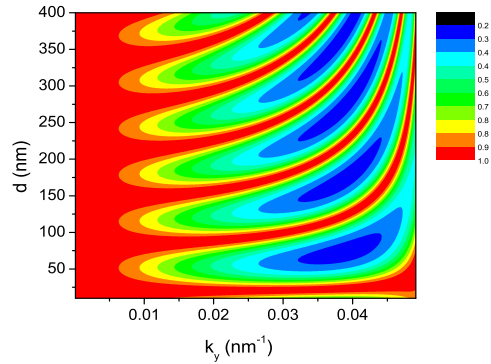}
\vspace*{-0.7cm} \caption{($d,k_y$) contour plot of the
transmission through a double barrier for  $U_0 = 100$ meV, $E=27$
meV, and $L=50$ nm.}
\end{figure}
\begin{figure}
\vspace{-1.2cm} \hspace*{-1cm}
\includegraphics*[height=6cm, width=8cm]{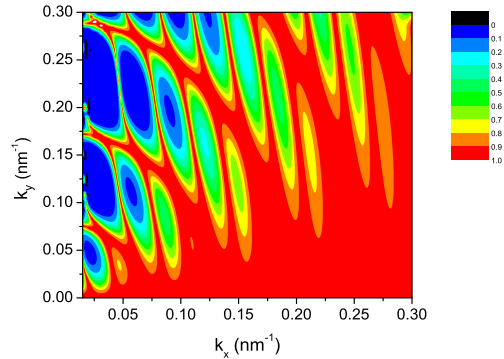}
\vspace*{-0.8cm} \caption{As in Fig. 1 but for a double well
with\\ energy $E
> U_0$, $U_0 = 50$ meV, and $L = 200$ nm. }
\end{figure}

\begin{figure}
\vspace{-0.8cm} \hspace*{-1cm}
\includegraphics*[height=5cm, width=7cm]{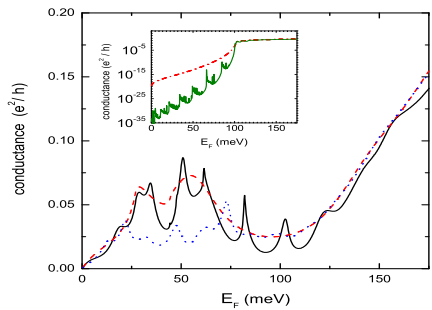}
\vspace*{-0.6cm} \caption{Conductance through  a single and double
barrier vs energy $E$ for $\Delta =0$. Inset:  conductance
through  the same structures for $m
v_F^2 = 270$ meV: 
the red (green) curve is for a single (double) barrier.}
\end{figure}

\begin{figure}
\vspace{-0.8cm} \hspace*{-1cm}
\includegraphics*[height=5cm, width=7cm]{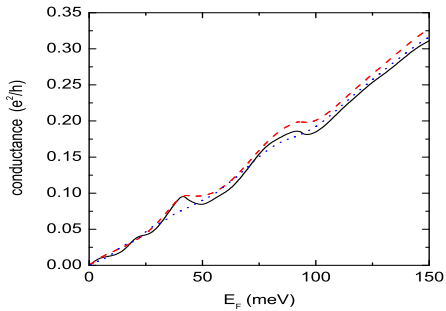}
\vspace*{-0.6cm} \caption{Conductance through  a single and double
well vs Fermi energy $E_{F}$. The parameters are given in the
text. }
\end{figure}

iii) {\it Tunneling current}. Selecting the wave vector components
$k_x$ and $k_y$ is possible using quantum point contacts. However,
experimentally one usually measures the "average" transmission.
That is, one usually measures the current $J$ which is
proportional to a weighted integral of the transmission $T(k_x,
k_y)$. Repeating the argument of Ref.
11 for 3D electrons with a
parabolic energy spectrum, we find that for the linear spectrum
$E=\hbar v_F k$ the current $J$, due to a voltage drop $eV$ along
the $x$ direction, is given by ($\lambda=2ev_F/h^2$)
\begin{equation}
J=-\lambda 
\int T(E,\theta) [f(E)-f(E+eV)] EdE \cos\theta d\theta,
\end{equation}
where $f(E)$ is the Fermi function. For low temperatures we can
approximate  $[f(E)-f(E+eV)]$ by  $-eV\delta (E-E_F)$ and extract
the low-bias conductance  from Eq. (4). For a 2D parabolic 
spectrum
$\lambda$ is replaced by  $2e\sqrt{2m}/h^2$ and $EdE$ by $E^{1/2}dE$ in Eq. (5). 
The main difference between the two cases is that in the latter
the transmission $T(E,\theta)$ depends only on the magnitude of
the $x$-component of the momentum, whereas in the former it
depends on both components.

In Fig. 5 we plot the conductance vs the Fermi energy for a
single-barrier (dashed curve) and for a double-barrier structure for
symmetric (solid curve) and asymmetric (dotted curve) barriers
with width
 $L = 50$ nm. The height of the first barrier is $U_0 = 100$ meV
and that of the second $U_0 =100, 50$, and $0$ meV, for the solid,
dashed, and dotted curve, respectively. The inter-barrier
separation is $d=50$ nm.
For energies below the maximum barrier height, the transmission is
dominated by the Klein tunneling effect, in which the incident
electrons are resonantly transmitted via the confined hole states
in the barriers, resulting in peaks in the conductance. An
additional
pronounced structure is also found for 
double barriers 
due
to an additional resonance effect caused by the quasi-confined
electron states in the well region. This resonant structure is
most explicit for  symmetric barriers. One striking consequence
of that is the fact that the low-bias conductance can be
significant {\it lowered} as the height of one of the barriers is
{\it decreased}. As the Fermi energy increases above the maximum
barrier height (here $100$ meV 
) the resonant
features practically disappear.
In contrast with the $\Delta\neq 0$ results shown in the inset,
which are a good qualitative approximation of the parabolic
spectrum case, the results for graphene,
 in the limits  $E_F
\rightarrow 0$ and $E_F \rightarrow \infty$, show $T=1$ for
near-normal incidence, see, e.g., Fig. 1. Using Eq. (4) we find
immediately $G \propto E_F$ which agrees with our numerical
results of Fig. 5. The present results are  valid only in the
ballistic regime, but in the presence of disorder the results for
the conductance near the Dirac point is expected to be strongly
modified.

In Fig. 6 we plot the conductance vs the Fermi energy for a single
well (dashed curve) and for a double well, symmetric (solid curve)
and asymmetric (dotted curve). The well width is $50$ nm. The
depth of the first well is $U_0 = 100$ meV and that of the second
$U_0 =100$, $50$, and $0$ meV, for the solid, dashed, and dotted
curve, respectively. The inter-well separation is $d=50$ nm. In
this case, only weak resonant features are found that are
determined completely by the first well. The presence of the
second well almost does not influence the conductance.

In summary, we reported unusual resonant-tunneling features in
graphene microstructures and a strong directional character of the
transmission through them.  The latter is perfect for a range of
incident angles. No such behavior is found in the case of a
parabolic spectrum. The conductance displays several resonant
features that can be tuned by the barrier heights as well as by
the separation between the barriers.

{\it Acknowledgements}
 This work was supported by the Brazilian
Council for Research (CNPq), the Flemish Science Foundation
(FWO-Vl), the Belgian Science Policy (IUAP) and the Canadian NSERC
Grant No. OGP0121756.

\end{document}